\documentstyle[amssymb,preprint,aps]{revtex}

\tightenlines

\begin{document}
\title{Inhomogeneous magnetism induced in a superconductor\\
at superconductor-ferromagnet interface}
\author{V. N. Krivoruchko$^{\ast }$ and E. A. Koshina}
\address{Donetsk Physics \& Technology Institute, Donetsk , Ukraine\\
(* Email address: krivoruc@host.dipt.donetsk.ua)}
\date{\today}
\maketitle

\section{Introduction}

Ferromagnetism and superconductivity are antagonistic to each other states
and ferromagnetism, being usually much stronger than superconductivity,
destroys the latter. However, when dealing with superconductor--ferromagnet (%
$SF$) hybrid structures the coexistence of these two phenomena is possible.
If a normal ferromagnetic metal ($F$) is in good electric contact with a
superconductor ($S$), the former can acquire some superconducting
properties, however, with spatially inhomogeneous order parameter [1,2]. The
spatial variation of the superconducting order parameter in the ferromagnet
arises as a response of the Cooper pair to the energy difference between the
two spin directions in ferromagnet. Due to nonzero momentum of the pairs,
the usually exponentially decaying Cooper's pair density in the $F$ layer
has a damped-oscillatory behavior (see, e.g., Ref. [3] for details).\ As has
been theoretically predicted, this causes the formation of the so-called $%
\pi $-type superconductivity [4,5], an enhancement of $dc$ Josephson current
in $SFIFS$ tunnel junctions [6,7] (where $I$ is an insulator) and the
transition temperature oscillation for $SF$ multilayers [8,9] as a function
of the exchange field $h_{F}$ or of the thickness $d_{F}$ of a ferromagnet.
Experiments that have been performed on $SFS$ weak links [10,11] and $SIFS$\
tunnel junctions [12]\ directly prove the $\pi $-phase superconductivity.
Planar tunneling spectroscopy reveals also a $\pi $-phase shift in the order
parameter, when superconducting correlations coexist with ferromagnetic
order [13]. However, experiments on $T_{C}(d_{F})$ $SF$ multilayers behavior
are still not so conclusive: \ the nonmonotonic oscillation-like behavior of 
$T_{C}(d_{F})$\ was also observed in ,e.g.,\ [14 -16] while negative results
were reported in, e.g., [17,18]. For interpretation of experimental results,
along with a mechanism of $\pi $-type superconductivity and the suppression
of $T_{C}$ due to a strong exchange field of a ferromagnet other mechanisms
were suggested such as a ''magnetically dead'' interfacial layer [16], the
effects of a finite interface transparency [18], spin-flip scattering [3,19]
and other.

The proximity induced superconductivity of the $F$ metal in $SF$ sandwiches
has been intensively studied, while not so much attention has been payed to
the modification of the electron spectrum of the superconductor in a region
near the $SF$ interface\ (an exception is a nonequlibrium case, that we do
not consider here). In many cases, the Curie temperature of the ferromagnet
is by order of magnitude larger than the superconducting critical
temperature $T_{C}$ , and, in a rough approximation, for $SF$ hybrid
structures the ''leakage of magnetism'' into the $S$ layer should be quite
strong. Indeed, let us consider the proximity effects at the $SF$ bilayer
interface.$\ $When diffusing into the ferromagnetic layer a superconducting
electron is subject to an interaction from the local exchange field. E.g.,
an electron with a spin, e.g. ''up'', has an extra energy $H_{exc}=\mu
_{B}h_{F}$ caused by the intrinsic magnetic field $h_{F}$ in ferromagnet ($%
\mu _{B}$ is the Bohr magneton). After tunneling back from the $F$ to $S$
layer, one can expect a kind of tunneling interaction that is similar to the
superconducting proximity effect. Actually, in the superconductor the
electron quickly loses its extra energy $\delta E$ $\simeq H_{exc}$ during
the time $\tau \symbol{126}\hslash /H_{exc}$ being related to a small range $%
\lambda _{F}\symbol{126}\hslash v_{F}/H_{exc}$ ($\hslash $ is Planck's
constant, $v_{F}$ is Fermi velocity). Such an equilibrium process leads to a
modification of the electron spectrum of the superconductor on nanoscale
length $\lambda _{F}$ . In most cases $\lambda _{F}\symbol{126}10^{-8}m$, so
one can assume the importance of such processes for hybrid
superconductor-ferromagnetic systems with thin $S$ layers. However, in spite
of its importance, a theoretical analysis of induced magnetic properties of
the $S$ metal being in contact with the $F$ one is still lacking.

The investigation of a ''magnetic proximity''effect in $SF$ nanostructures
is the purpose of our work. On the basis of the microscopic theory for
proximity coupled $SF$\ bilayer (Sec. II), we calculate the spatial
variation of the superconducting order parameter into the $S$ layer near the
interface as a function of the ferromagnetic exchange field and boundary
parameters (Sec. III A). Quasiparticle's densities of states (DOS) variation
in the $S$ layer as a function of the distance from the $SF$ boundary is
also considered on the same basis (Sec. III B). It is shown that the
superconducting pair subjected to the proximity induced exchange field
breaks, which leads to the formation of quasiparticle subbands in the
superconducting energy gap. In the absence of spin-orbit (spin-flip)
scattering, the subgap bands accommodate only one spin of quasiparticle.
Equilibrium local magnetization of the $S$ layer due to the induced exchange
correlation is also calculated (Sec. III C). We end with the Conclusion
where we emphasize the significance of the leakage of magnetism into
superconductor for experiments on $SF$ structures.

\section{Proximity effect model}

Let us consider the proximity effects in the $SF$ bilayer of a massive $S$ ($%
d_{S}\gg \xi _{S}\gg l_{S}$) and a thin $F$ ($l_{F}$ $\ll d_{F}\ll \xi _{F}$%
) layers, for the case of an arbitrary transparency of the $SF$ boundary,
the proximity effect strength and general relation between Curie temperature
of the $F$ metal and superconducting critical temperature of the $S$ metal.
Here $\xi _{S}=\left( \frac{D_{S}}{2\pi T_{C}}\right) ^{1/2}$and $\xi
_{F}=\left( \frac{D_{F}}{2H_{exc}}\right) ^{1/2}$ are the effective
coherence lengths; $D_{S,F}$ are the diffusion coefficients, $d_{S,F}$ are
the thicknesses and $l_{S,F}$ are the electron mean free paths of the $S$
and $F$ layers, respectively; $H_{exc}$ is the exchange energy of the
ferromagnet. (Henceforth, we have taken the system of units with $\hbar
=k_{B}=1$.) We assume the ''dirty'' limit for both metals. The
superconducting critical temperature of the $F$ material equals zero. All
quantities for $S$ layer are assumed to depend only on a single coordinate $%
{\it x}$ normal to the interface surface of the materials, while, due to\
the relation $d_{F}\ll \xi _{F}$ on thickness, one can neglect the
coordinate dependence within the $F$ layer. We also expect the $F$ layers
magnetization to be aligned parallel to the interface, so that they do not
create a spontaneous magnetic flux penetrating into the $S$ layer. Under
these conditions, the only magnetic interaction which can affect the
superconductor is the short-range exchange interaction between the
superconducting quasiparticles and magnetic moments into the ferromagnet.

As is well known, the normal and anomalous Green's functions of the
''dirty'' metals are described by the Usadel equations [20]. We will
restrict ourselves to the case when the spin-orbit scattering is absent and
spin ''up'' and spin ''down'' electron subbands do not mix with each other
(see, e.g., [3]). With the assumptions given above the $SF$ bilayer is
described by the following equations system for one (e.g.''up'') spin
subband (the domain ${\it x}\geq 0$ is occupied by the $S$ metal and ${\it x}%
<0$ by the $F$ metal):

\begin{equation}
\Phi _{S}=\Delta _{S}+\xi _{S}{}^{2}\frac{\pi T_{C}}{\omega G_{S}}%
[G_{S}{}^{2}\Phi _{S}^{\prime }]^{\prime },\text{ \ \ \ \ \ }G_{S}=\frac{%
\omega }{(\omega ^{2}+\Phi _{S}\tilde{\Phi}_{S})^{1/2}},  \label{ref1}
\end{equation}

\begin{equation}  \label{ref2}
\Delta_S\ln(T/T_C)+2\pi T \sum_{\omega>0}[(\Delta_S-\Phi_S G_S) /\omega]=0,
\end{equation}

\begin{equation}
\Phi _{F}=\xi _{F}{}^{2}\frac{\pi T_{C}}{\tilde{\omega}G_{F}}%
[G_{F}{}^{2}\Phi _{F}^{\prime }]^{\prime },\text{ \ \ \ \ \ \ }G_{F}=\frac{%
\tilde{\omega}}{(\tilde{\omega}^{2}+\Phi _{F}\tilde{\Phi}_{F})^{1/2}}.
\label{ref3}
\end{equation}
Here $\Delta _{S}$ is the superconducting order parameter of the $S$
material, $\tilde{\omega}=\omega +iH_{exc}$, where $\omega \equiv \omega
_{n}=\pi T(2n+1)$, $\ n=\pm 1,\pm 2,\pm 3,...$ is Matsubara frequency; the
summation over frequencies in (2) is cut off by the Debye frequency $\omega
_{D}$ ; $\tilde{\Phi}(\omega ,H_{exc})=\Phi ^{\ast }(\omega ,-H_{exc})$; the
prime denotes differentiation with respect to a coordinate ${\it x}$. (Note
that the momentum renormalization in the ferromagnet is not so important as
the frequency renormalization.) We introduce the modified Usadel functions $%
\Phi _{S}=\omega F_{S}/G_{S}$, $\Phi _{F}=\tilde{\omega}F_{F}/G_{F}$ , where 
$G_{F,S}$ and $F_{F,S}$ are Green's functions for the $F$ and $S$ material,
respectively, to take into account the usual normalized confinement on the
Green's functions $G_{F,S}$ and $F_{F,S}$ (see Refs. [21,22] for details).

The Eqs. (1)-(3) should be supplemented with the usual boundary conditions
in the bulk of the $S$ metal: $\Phi _{S}(\infty )=\Delta _{S}(\infty
)=\Delta _{0}(T)$ , where $\Delta _{0}(T)$ is the $BCS$ value of the order
parameter, and at the external surface of the $F$ metal $\Phi
_{F}^{/}(-d_{F})=0$\ . The boundary conditions at the $SF$ interface are
[22]:

\begin{center}
\begin{equation}
\left. \frac{1}{\tilde{\omega}}\gamma \xi G_{F}{}^{2}\Phi _{F}^{\prime
}\right| _{x=0}=\left. \frac{1}{\omega }\xi _{S}G_{S}{}^{2}\Phi _{S}^{\prime
}\right| _{x=0},  \label{ref4}
\end{equation}
\end{center}

\begin{equation}
\left. \xi \gamma _{BF}G_{F}\Phi _{F}^{\prime }\right| _{x=0}=\left. \tilde{%
\omega}G_{S}(\Phi _{S}/\omega -\Phi _{F}/\tilde{\omega})\right| _{x=0},
\end{equation}
where $\gamma =\rho _{S}\xi _{S}/\rho _{F}\xi $ is a measure of the
proximity effect strength, $\gamma _{BF}=R_{B}/\rho _{F}\xi $ describes the
effect of the boundary transparency. Here $\rho _{S,F}$ are the normal-state
resistivities of the $S$ and $F$ metals, $R_{B}$ is the product of the $SF$
boundary resistance and its area. Here and below we write our formulas for
the $F$ metal using the effective coherence length of normal nonmagnetic ($N$%
) metal $\xi =\left( \frac{D_{F}}{2\pi T_{C}}\right) ^{1/2}$ with the
diffusion coefficient $D_{F}$\ instead of \ $\xi _{F}=\left( \frac{D_{F}}{%
2H_{exc}}\right) ^{1/2}$, to have a possibility to analyze both limits $%
H_{exc}\rightarrow 0$ ($SN$ bilayer with $d_{N}\ll \xi $) and $H_{exc}>>\pi
T_{C}$\ . The relations (1)-(5) generalize the proximity effect problem with
an arbitrary interface transparency for the case of normal metal with
ferromagnetic order.

\section{Results}

\subsection{The spatial variation of the order parameter}

Due to a mesoscopic thickness of the $F$ metal the proximity effect problem
can be reduced to the boundary problem for the $S$ layer and a relation for
determining $\Phi _{F}$ [22,23]. There are three parameters which enter the
model: $\gamma _{M}=\gamma d_{F}/\xi $ is the measure of the strength of
proximity effect between the $S$ and $F$ metals, $\gamma _{B}=\gamma
_{BF}d_{F}/\xi $ describes the electrical quality of the $S/F$ boundary, and 
$H_{exc}$ is the energy of the exchange correlation in the $F$ layer. In the
general case, the problem needs self-consistent numerical solution of the
Usadel equations (such mode consideration see, e.g., in Ref. [24]). Here,
however, we will not discuss the quantitative calculations, but present
qualitatively correct ones to consider new physics we are interested in. In
this approximation, for bilayer with weak proximity effect $\gamma _{M}\ll 1$
and arbitrary value of the boundary resistance $\gamma _{B}$ (quite a
realistic experimental case) for the function $\Phi _{S}$ we have the
analytical solution (see Ref. [22,23]):

\begin{equation}
\Phi _{S}(\omega ,x)=\Delta _{0}\{1-\gamma _{M}\beta \tilde{\omega}\frac{%
\exp (-\beta x/\xi _{S})}{\gamma _{M}\beta \tilde{\omega}+\omega A(\omega )}%
\}
\end{equation}

where $\beta ^{2}=\left( \omega ^{2}+\Delta _{0}^{2}\right) ^{1/2}/\pi
T_{C}, $ and\ $A(\omega )=\left[ 1+\gamma _{B}\tilde{\omega}\left( \gamma
_{B}\tilde{\omega}+2\omega /\beta ^{2}\right) /(\pi T_{C})^{2}\right] ^{1/2}$%
. Using $\Phi _{S}(\omega ,x)$ (6) and the self-consistency condition (2)
one can obtain the spatial variation of the order parameter in the $S$ layer 
$\Delta _{S}(x)$ for different values of $\gamma _{B}$ and $\gamma _{M}\ll 1$%
. As follows from the analysis, the $BCS$ value of the order parameter is
reached at the distance of several $\xi _{S}$ into the $S$ layer. At fixed
value of the exchange interaction $H_{exc}$ , the order parameter at $SF$
boundary $\Delta _{S}(0,H_{exc})$ decreases with increasing boundary
transparency, which leads to a decrease in the jump of the amplitude of the
Cooper pairs in going from the $S$ to the $F$ layer. A decrease in the
boundary transparency leads to a sharp increase of the order parameter at $%
SF $ boundary. The behavior on $\gamma _{M}$ and $\gamma _{B}$ parameters
here is analogous to the situation in $\ $the $SN$ system (see, e.g., Ref.
[21]).

Another important feature of Eq.(6) is its dependence upon\ $H_{exc}$ . The
exchange interaction influence on the spatial variation of the order
parameter in the $S$ layer is shown on Fig. 1. Namely, the curves show the
dependence of difference of the order parameters in the $S$ layer for the
case, when magnetic interaction is turned off ($SN$\ bilayer) and the case
with ferromagnetic correlation ($SF$ bilayer) on distance from interface for
fixed boundary parameters and different strength of the $F$ layer exchange
field $H_{exc}:$\ $[\Delta _{S}(x,H_{exc}=0)-\Delta _{S}(x,H_{exc})]$\ . It
is seen that magnetic influence decreases with the increasing of the
distance from $SF$ boundary and at $x>2\xi _{S}$ the order parameter becomes
equal to those for the case of $SN$ bilayer. This is not surprising, since
one would expect that the induced exchange field should suppress the
superconducting order parameter at some distance into the $S$ layer in
excess of the $N$ layer suppression. Suppression increases with the increase
of the exchange interaction parameter\ $H_{exc}$\ and of the proximity
effect. The curves on Fig. 1 illustrate the spatial dependencies of\ the
induced exchange correlation in the superconductor for the case of vanishing
interface resistance $\gamma _{B}=0$ . With an increase of the $SF$ boundary
resistance the electrical coupling of the\ $S$\ and $F$\ metals decreases
and in the limit $\gamma _{B}\longrightarrow \infty $ the metals become
decoupled. As follows from (6), far ($x\gg \xi _{S}$ ) from the interface
the bulk superconductivity is re-established.

\subsection{Spin-splitting of DOS and local exchange-interaction induced
states}

Spin splitting of DOS in the $S$ layer is the second demonstration of a
magnetism leakage\ into the superconductor. The Green's functions for the $S$
layer $G_{S\uparrow \uparrow }(\omega ,x)$ and $G_{S\downarrow \downarrow
}(\omega ,x)$ for both spin subbands one can obtain using the solution (6)
for the functions $\Phi _{S}(\omega ,x)$ with $\tilde{\omega}=\omega
+iH_{exc}$\ and $\tilde{\omega}=\omega -iH_{exc}$\ , respectively.
Performing the analytical continuation to the complex plane by the
substitution $\omega \rightarrow -i\varepsilon $ we calculate the spatial
dependence of quasiparticle DOS for each spin subband $N_{S\uparrow
}(\varepsilon ,x)=ReG_{S\uparrow \uparrow }(\omega ,x)$ and $N_{S\downarrow
}(\varepsilon ,x)=ReG_{S\downarrow \downarrow }(\omega ,x)$. On Fig. 2 the $%
N_{S\uparrow }(\varepsilon ,x)$ dependence at different distances from the $%
SF$ interface is presented for $T<<T_{C}$, $H_{exc}=5\pi T_{C}$ and $\gamma
_{M}$ = 0.1, and vanishing boundary resistance ($\gamma _{B}=0)$ . We find
that for the case of $H_{exc}\neq 0$, and $\gamma _{M}\neq 0$ the
quasiparticle DOS is spin-splitted, that means $N_{S\uparrow }(\varepsilon
,x)\neq $ $N_{S\downarrow }(\varepsilon ,x)$ and $N_{S\uparrow }(\varepsilon
,x)\neq $ $N_{S\uparrow }(-\varepsilon ,x)$, in the $S$ layer at the
distance of a few $\xi _{S\text{ }}$from the $SF$ boundary. As one can
expect from the fermionic symmetry, the spin-up particles and spin-down
holes have the same DOS, and likewise for spin-down particles and spin-up
holes. The spin-splitting decreases with an increase of the distance from
the boundary and vanishes in the bulk of the$\ S$ layer (see curve 4 on Fig.
2). The DOS of Cooper's pairs $\aleph _{\uparrow }(\varepsilon
,x)=ReF_{S\uparrow \downarrow }(\omega ,x)$, $\aleph _{\downarrow
}(\varepsilon ,x)=ReF_{S\downarrow \uparrow }(\omega ,x)$ is spin-splitted,
too. These features are due to the initial spin-splitting electrons on the
Fermi surface in the $F$ metal and characterize the $SF$ bilayer as a united
system.

On Fig. 3 the $N_{S\uparrow }(\varepsilon ,x)$ dependence is presented for $%
x/\xi _{S}=1$ and different values of the exchange energy for bilayer with
finite boundary transparency $\gamma _{B}=0.1$ and proximity effect strength 
$\gamma _{M}=0.1$. The behavior on $\gamma _{B}$ parameter is due to two
mechanisms: with decreasing boundary transparency spin-splitting increases
if $\gamma _{B}\eqslantless 0.2$, however, for large enough $\gamma _{B}>0.2$%
\ the decrease of the interface electrical quality preponderates and DOS
approaches the BCS value.

Other important features, shown on Figs. 2 and 3, are the local states that
exist inside the energy gap at the distances of a few $\xi _{S}$ from the $%
SF $ boundary. These subgap states are absent at the $SF$ boundary ($x=0$),
far from the $SF$\ interface, and also absent if $H_{exc}=0$. For small
values of $\gamma _{M}$ and $\gamma _{B}=0$ , as follows from Eq.(2) and
solution (6), $N_{S\uparrow }(\varepsilon ,x)$ has singularities for

\begin{equation}
\varepsilon =\pm \Delta _{0}\{1-\frac{\gamma _{M}\beta _{\varepsilon }\tilde{%
\varepsilon}}{\varepsilon +\gamma _{M}\beta _{\varepsilon }\tilde{\varepsilon%
}}\exp (-\beta _{\varepsilon }x/\xi _{S})\}
\end{equation}

where $\beta _{\varepsilon }^{2}=(\Delta _{0}^{2}-\varepsilon
^{2})^{1/2}/\pi T_{C}$ and$\ \tilde{\varepsilon}=\varepsilon -H_{exc}$. We
found the singularities inside the superconducting gap, $-\Delta
_{0}<\varepsilon <\Delta _{0}$ , by numerical calculations. The solutions of
Eq.7 for $x/\xi _{S}=0,1,5$\ and for $\gamma _{M}=0.1$, $H_{exc}=5\pi T_{C}$
are presented on Fig.4 in the diagram form. The local states are thus
definitely not due to the spatial variation of the pair potential, but due
to Cooper pairs breaking in the superconductor by the exchange-induced
magnetic correlation. The region of their existence increases with the
increasing of $H_{exc}$, or increasing pair breaking effects. In the absence
of spin-flip (e.g., spin-orbit) scattering, the subgap bands accommodate
only one spin of quasiparticles. These bands bear superficial resemblance to
both the bands observed at interface of superconductor and perfectly
insulating ferromagnet [25] and bulk superconductor containing finite
concentrations of magnetic impurities [26,27].

\subsection{Local magnetization}

As was mentioned above, the influence of the ferromagnet on the
superconductor is reflected in a nonzero value of the difference in the
densities of states for spin-up and spin-down electrons, $N_{S\uparrow
}(\varepsilon ,x)$ and $N_{S\downarrow }(\varepsilon ,x)$\ . This DOS
splitting causes inhomogeneous quasiparticle spin density in the $S$ metal,
i.e., an effective magnetization $M_{S}(x)$ of the $S$ layer, that can be
found using the relation (see, e.g., [28]) :

\begin{equation}
M_{S}(x)=M_{O}\int_{0}^{\infty }d\varepsilon \{N_{S\uparrow }(\varepsilon
,x)-N_{S\downarrow }(\varepsilon ,x)\}f(\varepsilon )
\end{equation}

where $M_{O}=gS_{e}\mu _{B}(=\mu _{B})$\ ,\ $S_{e}=1/2,g=2$ and $%
f(\varepsilon )=1/[\exp (\varepsilon /T)+1]$ is Fermi distribution function.
This is confirmed by numerical calculations of $M_{S}(x)$ Eq.(8) shown in
Figs. 5-7. Fig. 5 shows the leakage of magnetization into the superconductor
versus distance from the $SF$ interface for fixed boundary parameters and
different exchange field energy values. Fig. 6 shows the same magnetic
characteristics but for $SF$\ sandwich with fixed exchange energy and
boundary transparency, and different proximity effect strength.\ The curves
on Fig. 7 illustrate inhomogeneous magnetism induced into the superconductor
by proximity effect, $M_{S}(x)$ , for the $SF$ bilayers with different
interface transparency and fixed $H_{exc}$ and $\gamma _{M}$ . Following DOS
dependance, the magnetization behavior\ on $\gamma _{B}$ parameter is due to
two mechanisms: firstly, $\gamma _{B}\eqslantless 0.2$ , with decreasing
boundary transparency spin-splitting increases and $M_{S}(x)$ increases too,
secondly however, for large enough $\gamma _{B}>0.2$\ the decreasing of the
interface electrical quality preponderates and induced magnetization
decreases.

All the curves of $M_{S}(x)$ show essentially non-linear behavior versus
distance from the $SF$\ interface, and have important common feature - a
damped oscillatory behavior on a scale of superconducting coherence length $%
\xi _{S}$ . As is evident from the plots on Figs. 5-7 , induced by proximity
effect magnetic correlations in the $S$ layer are inhomogeneous in character
with spatially dependent magnitude and direction of its molecular field.

\section{Conclusion}

We have calculated the magnetic proximity effect induced in the
superconductor at $SF$ interface. We find that an equilibrium exchange of
electrons between the $F$ and $S$ metals results not only in proximity
induced inhomogeneous superconductivity of the $F$ metal, as was found
earlier, but in proximity induced inhomogeneous magnetism of the $S$ metal,
too. Our results show that the appearing damping magnetic correlations in
the superconductor are quite robust and, to our opinion, are quite possible
for experimental measurements.

Experimental studies of magnetic proximity effect in $SF$ structures are
nowadays unclear and controversial. Tedrow {\it et. al}. [29] were the first
who carried out high-field tunneling measurements of the quasiparticle DOS
in superconducting films of $Al$ backed by $EuO$, which in bulk form is a
ferromagnetic insulator with a Curie temperature $T_{CR}=70K$. The tunnel
conductances of these junctions show a spin splitting of the $Al$ DOS in
excess of the splitting observed for $Al$ films in contacts with nonmagnetic
substrates in equivalent magnetic field. The authors interpreted the excess
spin splitting in terms of a coupling between the $Al$ conduction electrons
and the $EuO$ magnetization. However, Tokuyasu {\it et.al}. [30] argued that
the observed excess Zeeman splitting in $EuO/Al$ proximity contacts cannot
be explained in such a mode, and considered that excess splitting is the
reflection of the exchange coupling of the tunneling electrons with $Eu^{2+}$
spins at the interface. Recent experiments [31,15] revealed that the
effective thickness of the magnetic layer in $SF$ hybrid structures usually
is much larger than its physical thickness. The increase of the thicknesses
is so great that in all samples, except for those with extremely thin
magnetic layers, the crossover to a 3D state superconductivity is never in
fact observed experimentally. This is to be contrasted with the case of
nonmagnetic spacer layers, where these two length scales are comparable.
Taking into account our results, we explain that the rise of the effective
magnetic layer thickness in the $SF$ sandwiches is due to magnetic proximity
effect. Namely, the induced magnetic correlation into the $S$ layer depletes
Cooper's pair density at the $SF$ boundary that results in an excess
thickness of the magnetic layer.

In conclusion, we study the magnetic correlations into a superconductor that
have been induced at $SF$ interface due to proximity effect. Equilibrium
leakage of magnetism into the $S$ metal results in (i) additional spatial
suppression of the order parameter in the $S$ layer, (ii) spatial spin
splitting of the quasipaticles DOS, (iii) local bands that appear inside the
energy gap in the $S$ layer, and (iv) spatially dependent magnitude and
direction induced magnetization of the $S$ layer. The existence of these
magnetic properties of the $S$ metal is quite important for $SF$
nanostructures and should be taken into account while comparing theoretical
results with experimental data.

\section{Acknowledgment}

The authors are grateful to V. V. Ryazanov for useful discussions.

\begin{center}
Figure Captures
\end{center}

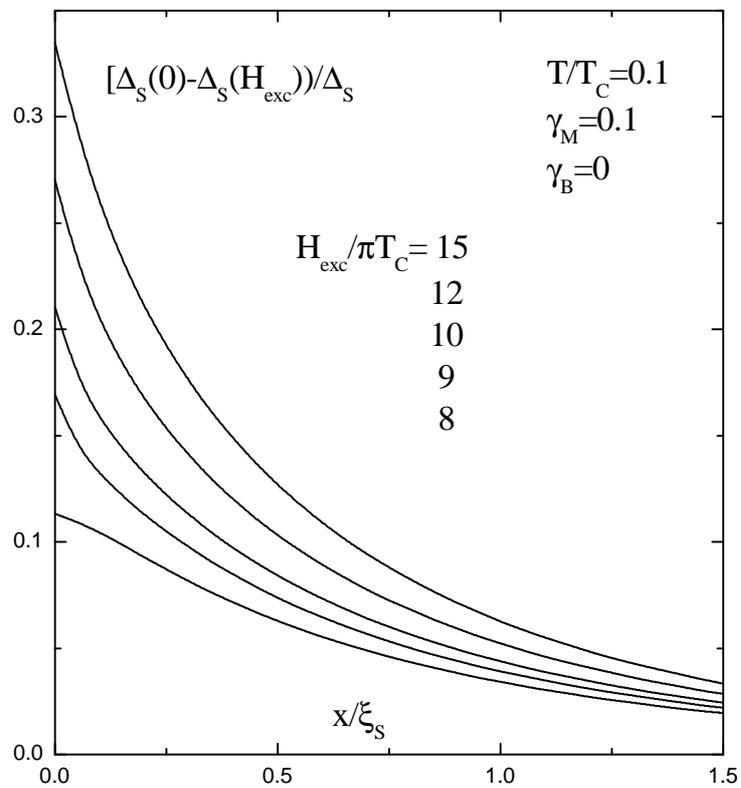
Fig. 1. The difference of the superconducting order parameter in the $S$
layer versus distance from the interface for $SN$ and $SF$ structures with
the same boundary parameters ($\gamma _{M}=0.1$ , $\gamma _{B}=0$ ), and
different ferromagnetic field energy $H_{exc}/\pi T_{C}=$8, 9, 10, 12 and 15.

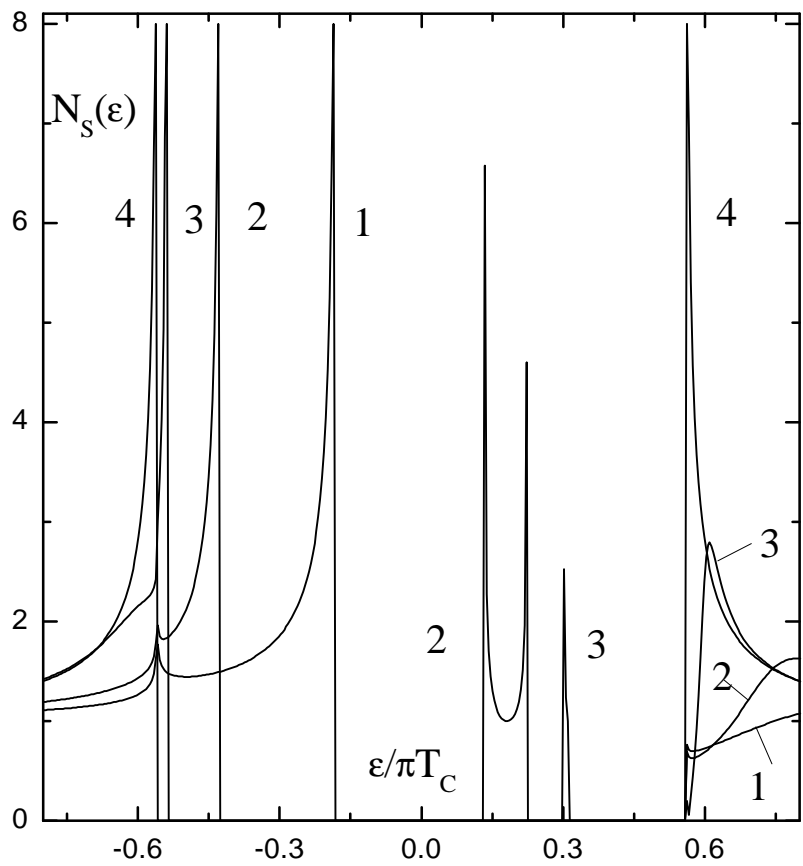
Fig. 2. Normalized density of state for spin ''up'' quasiparticles in the $S$
layer of the $SF$ sandwich for $\gamma _{M}=0.1$ , $\gamma _{B}=0$ and $%
H_{exc}=5\pi T_{C}$ , and various distances from the $SF$ interface: $x/\xi
_{S}=$ 0, 1, 5, and 30 (curves 1, 2, 3, and 4, respectively).

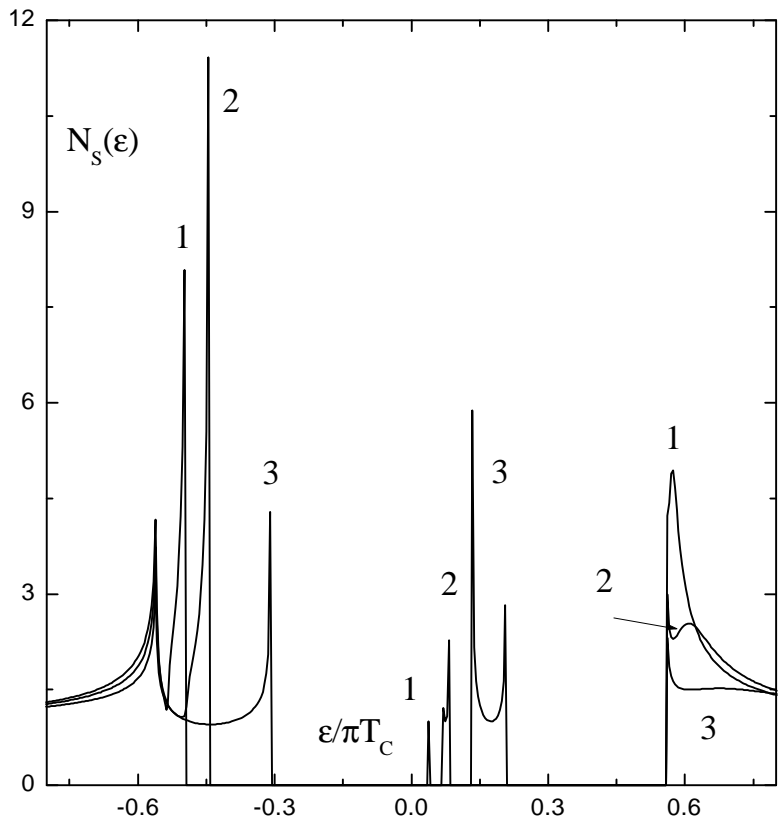
Fig. 3. Same as on Fig. 2 for $\gamma _{M}=0.1$ , $\gamma _{B}=0.1$ and $%
x=\xi _{S}$ , and various ferromagnetic field energies: $H_{exc}/\pi T_{C}=$%
1, 2, and 5 (curves 1, 2, and 3, respectively).

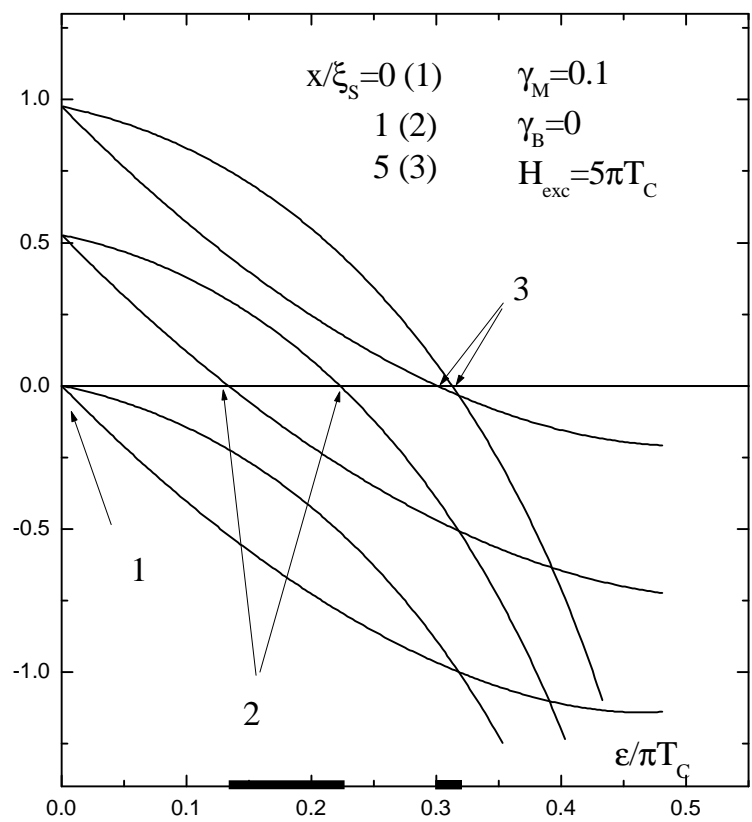
Fig. 4. Diagram solutions of the Eq.7 for $\gamma _{M}=0.1$, $H_{exc}=5\pi
T_{C}$ and $x/\xi _{S}=$ 0, 1, 5, (curves 1, 2, and 3, respectively). The
regions where the local states exist are shaded; see also Fig. 2 for
comparison.

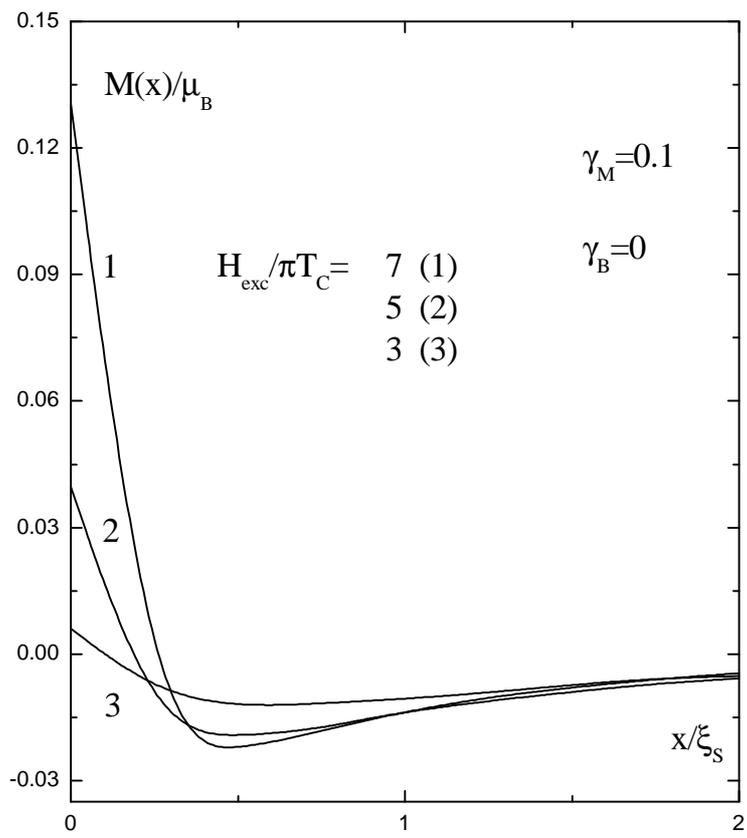
Fig. 5. Leakage of magnetization into the $S$ material versus distance from
the interface for $SF$ sandwich for $\gamma _{M}=0.1$ , $\gamma _{B}=0$ ,
and different exchange energies $H_{exc}/\pi T_{C}=$7, 5, and 3 (curves 1,
2, and 3, respectively).

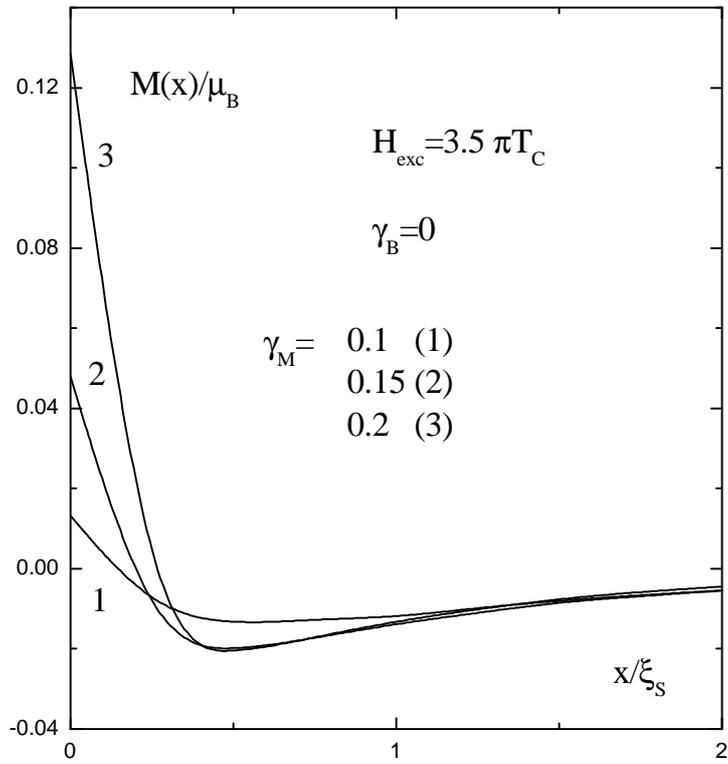
Fig. 6. Same as on Fig. 4 for $\gamma _{B}=0$ , $H_{exc}=3.5\pi T_{C}$ and
different proximity effect strength $\gamma _{M}$ =\ 0.1, 0.15, 0.2 \
(curves 1, 2, and 3, respectively).

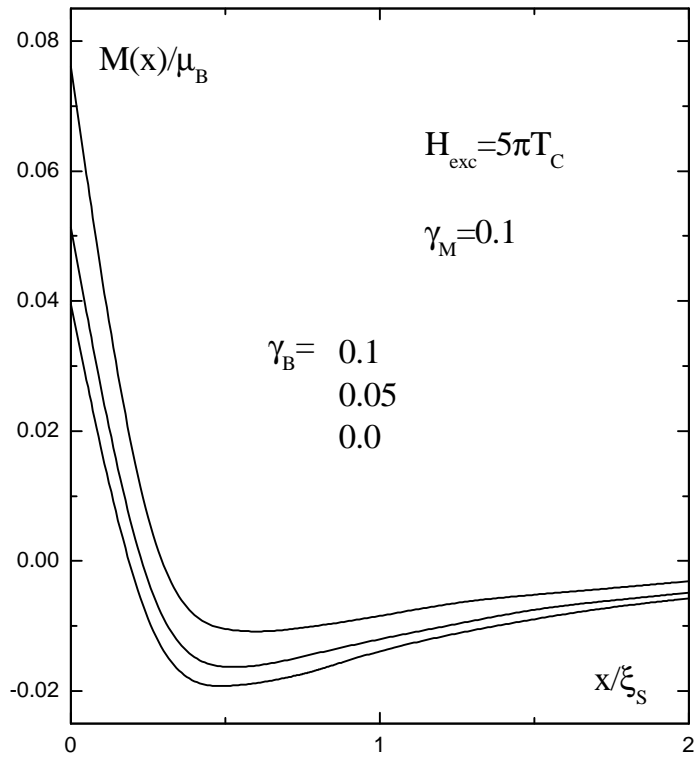
Fig. 7. Same as on Fig. 4 for $\gamma _{M}=0.1$ , $H_{exc}=5\pi T_{C}$ and
different $SF$ boundary transparency $\gamma _{B}$ =\ 0.0, 0.05, and 0.1\
(curves 1, 2, and 3, respectively).

\end{document}